# Retrieving the Aerosol Complex Refractive Index using PyMieScatt: A Mie Computational Package with Visualization Capabilities

Benjamin J. Sumlin[1], William R. Heinson[1], and Rajan K. Chakrabarty[1,2]*

[1] Center for Aerosol Science and Engineering

Department of Energy, Environmental and Chemical Engineering

Washington University in St. Louis, Missouri, USA – 63130

[2] McDonnell Center for the Space Sciences

Washington University in St. Louis, Missouri, USA – 63130

* Correspondence to: chakrabarty@wustl.edu

Abstract

The complex refractive index $m=n+ik$ of a particle is an intrinsic property which cannot be directly measured; it must be inferred from its extrinsic properties such as the scattering and absorption cross-sections. Bohren and Huffman called this approach "describing the dragon from its tracks", since the inversion of Lorenz-Mie theory equations is intractable without the use of computers. This article describes *PyMieScatt*, an open-source module for Python that contains functionality for solving the inverse problem for complex *m* using extensive optical and physical properties as input, and calculating regions where valid solutions may exist within the error bounds of laboratory measurements. Additionally, the module has comprehensive capabilities for studying homogeneous and coated single spheres, as well as ensembles of homogeneous spheres with user-defined size distributions, making it a complete tool for studying the optical behavior of spherical particles.

**Keywords:** Aerosol Optics; Mie Theory; Python 3; Electromagnetic Scattering and Absorption; Open-source Software

# 1 Introduction

Craig Bohren and Donald Huffman's (B&H [1]) FORTRAN codes for Lorenz-Mie (Mie hereafter) theory [2] (BHMIE) have been the foundation and inspiration for generations of computer codes applicable to aerosol optics, written in many languages for a variety of problems. However, to our knowledge, a comprehensive collection of Mie codes is not available for Python, nor are there any open-source codes for solving the inverse Mie problem to retrieve the complex refractive index ($m = n+ik$) for a particle of known size parameter. Python (http://www.python.org) is an open-source scripting language that is popular for scientific, mathematical, and engineering computing. To address the lack of Mie optics software for Python, we wrote PyMieScatt, or the <u>Py</u>thon <u>Mie</u> <u>Scatt</u>ering package. PyMieScatt is a module, similar to a library in C. These codes are meant to be a tool for researchers, and can be used as either a standalone calculator or to develop custom Python scripts for specialized research or educational purposes.

PyMieScatt evolved from efforts to translate Christian Mätzler's collection of MATLAB scripts [3] into an open-source, platform-independent language and expand them for common problems found in experimental aerosol optics. Where Mätzler's work is a translation of the original BHMIE code, PyMieScatt has been written to emphasize readability and the mathematics of Mie Theory. Many additional components for computational work relevant to contemporary aerosol optics have been added. The package can:

i. calculate Mie efficiencies for extinction ($Q_{ext}$), scattering ($Q_{sca}$), absorption ($Q_{abs}$), radiation pressure ($Q_{pr}$), and backscatter ($Q_{back}$), as well as the asymmetry parameter ($g$) for a single particle in the Mie and Rayleigh regimes;
ii. as an extension of (i), calculate coefficients ($\beta_{abs}$ and $\beta_{sca}$, for example) using measured, user-defined, or mathematically generated size distribution data for an ensemble of homogeneous particles;
iii. produce arrays for plotting the angular-dependent light scattering intensity for parallel and perpendicular polarizations in both $\theta$-space and q-space [4] for a single particle or an ensemble of particles;
iv. calculate the four nonzero scattering matrix elements $S_{11}$, $S_{12}$, $S_{33}$, and $S_{34}$ as functions of scattering angle;
v. do (i), (iii), and (iv) for coated spheres (core-shell particles);
vi. solve the inverse Mie problem for complex $m = n+ik$, given inputs of scattering, absorption, and size parameter for a single particle, or ensemble of particles given additional size distribution data and calculate solution regions bound by measurement uncertainty;
vii. as an extension of (vi), use additional measurements of backscatter efficiency ($Q_{back}$) or coefficient ($\beta_{back}$) to constrain the inverse problem to produce unique solutions.

Points (vi) and (vii) represent the major result of this work. While points (i) through (v) are useful for solving common problems, (vi) and (vii) use our development of a novel method for solving the inverse Mie problem for the complex refractive index when size parameter is known. Often, $m$ is the unknown parameter in experiments involving a particle or a distribution of particles. Mie equations take *m* as an input, and so the problem of deriving it from measured data represents an inverse problem that is inconvenient to solve without the use of computers. Aerosol optics experiments frequently involve instruments designed to measure light scattering and absorption, such as photoacoustic spectrometers [5] and integrating nephelometers [6] which directly measure absorption and scattering coefficients $β_{abs}$ and $β_{sca}$, respectively. Using Mie theory, one can calculate these parameters directly with knowledge of the analyte's size distribution and effective index of refraction (where it is known or assumed that the constituent particles are spherical). Conversely, with knowledge of optical behavior and morphology as measured by laboratory equipment, PyMieScatt can determine *m* by solving this inverse problem.

Simulation of scattering and absorption by an arbitrary particle is trivial with modern computing power and algorithms such as *T*-Matrix [7, 8], finite-difference time-domain [9], the discrete dipole approximation [10], or a Mie theory implementation of choice. However, the inverse problem, whether it is constructed to solve for particle size, real *m* or complex *m*, is confounded by the multidimensional parameter space the equations require. Without a direct imaging technique like tomography, the problem must be constrained by measuring the particle morphology, making certain assumptions about the system, or careful parameterization of variables. There are numerous examples in the literature demonstrating the use of inversion techniques, in particular a variety of medical applications to infer properties about the health and morphology of blood cells [11-14], or to characterize the shape of individual bacteria [15-17]. Previous work to invert optical measurements to reconstruct particle properties are often constrained to the particle size and real *m* [18, 19], or use measurements at multiple angles to determine *m* [11].

We constrain the inverse problem by measuring or assuming the size of a spherical, homogeneous particle (or size distribution of an ensemble thereof), and knowing the wavelength of light illuminating it. Our parameter space is therefore less complex and our method constructs an inverse problem using measurements of scattering and absorption, which can be further constrained by introducing backscattering efficiency, a third independent parameter.

## 3 The Inverse Mie Problem

PyMieScatt uses a highly visual, geometric method to invert the Mie problem and calculate *m* for a single particle (given $Q_{abs}$, $Q_{sca}$, the wavelength, and diameter) or for an ensemble of particles (given $β_{abs}$, $β_{sca}$, the wavelength, and a size distribution). This contour intersection method visualizes various optical parameter spaces as functions of *n* and *k* and looks for intersections in the curves defined by optical measurements. Additionally, we include a strictly numerical approach that borrows certain principles from the visual technique and minimizes errors in the retrieval by brute-force iterative methods.

### 3.1 The contour intersection approach

The contour intersection inversion method determines the index of refraction by first computing $Q_{sca}(n,k)$ and $Q_{abs}(n,k)$ for a given wavelength and particle diameter across a range of *n* and *k*, then identifying the contours corresponding to measured $Q_{sca}$ and $Q_{abs}$, and finally identifying their intersections in *n-k* space. Intersections represent values of *m* that would produce the desired $Q_{sca}$ and $Q_{abs}$. Similarly, this method can also find the effective *m* of a size distribution of spherical particles by computing $\beta_{sca}(n,k)$ and $\beta_{abs}(n,k)$ and following the same procedure.

Figure 1 illustrates the theory of this process with an example. With ideal laboratory equipment, assume a test particle whose diameter was measured to be 300 nm, and its 375 nm $Q_{sca}$ and $Q_{abs}$ were 1.315 and 1.544, respectively. First, calculate the $Q_{sca}$ and $Q_{abs}$ surfaces over a range of *n* and *k* (panel A, note that we have added contour lines to the surfaces and have plotted *k* on a logarithmic scale to reveal the complex surface geometry). Then, we identify the contours corresponding to $Q_{sca} = 1.315$ and $Q_{abs} = 1.544$ (panel B). Finally, we project those contours to *n-k* space and locate their intersections (panel C, where *k* is plotted on a linear scale).

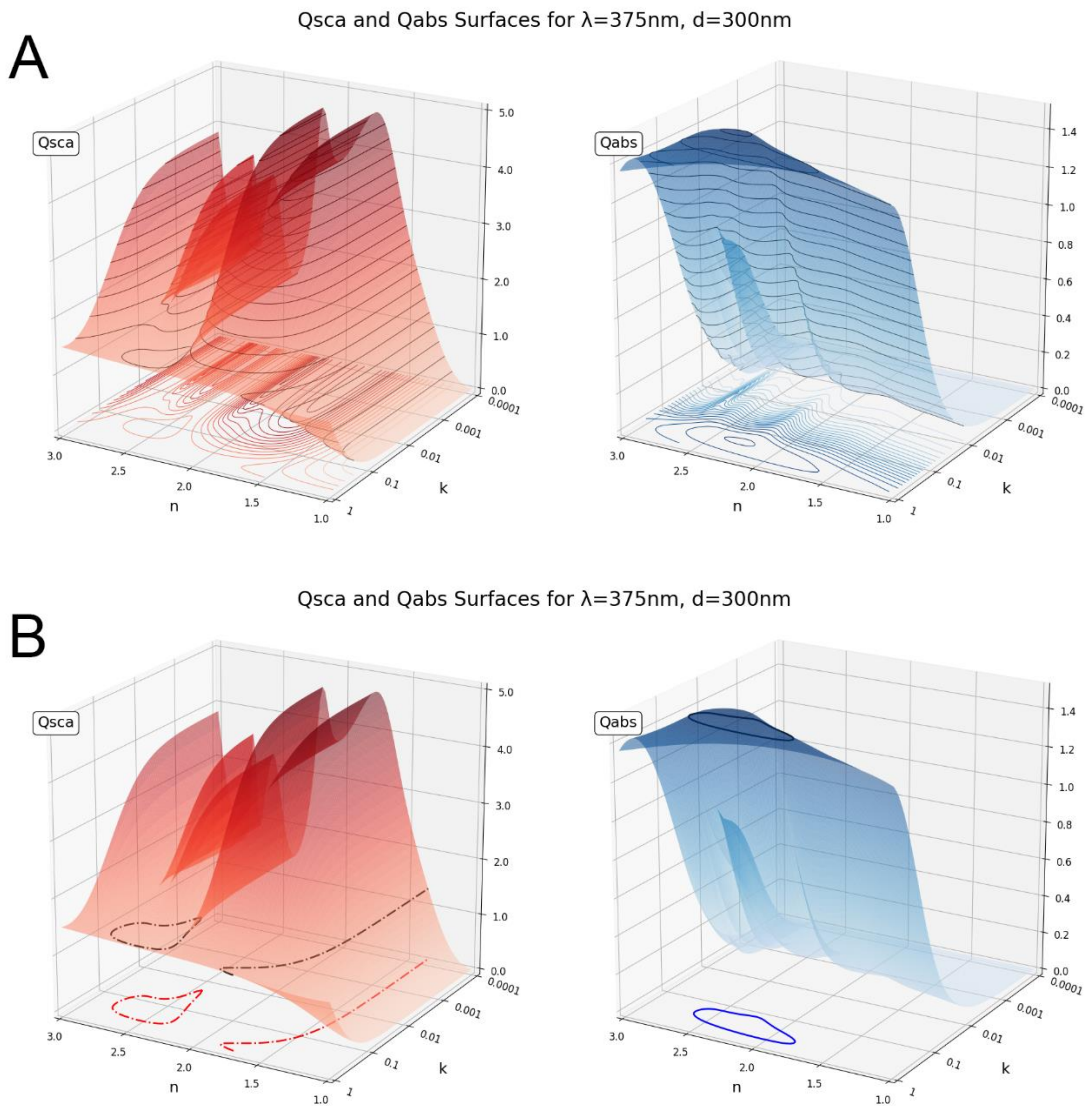

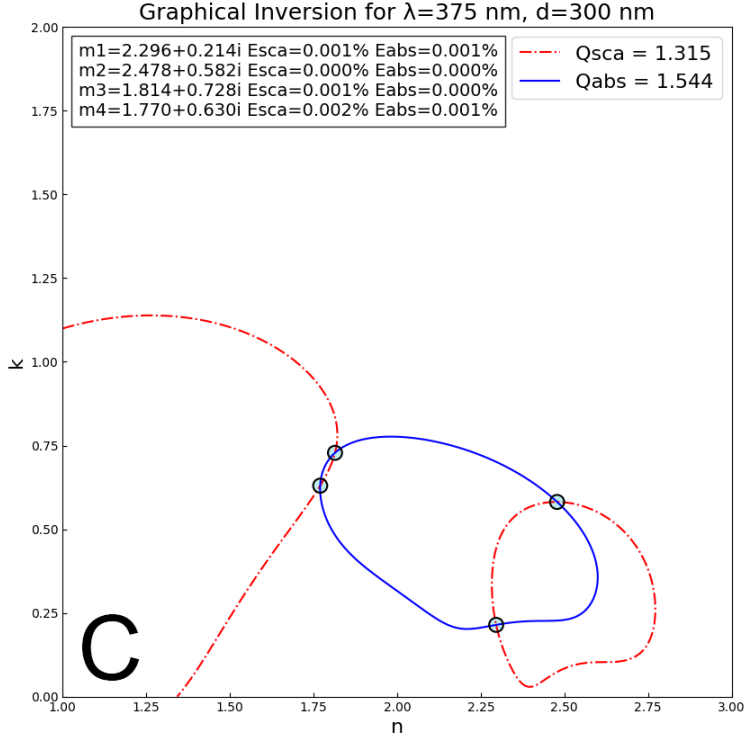

**Figure 1. A:** The $Q_{sca}$ (left, red) and $Q_{abs}$ (right, blue) surfaces for a range of *m* with λ = 375 nm and d = 300 nm. **B:** The contours where $Q_{sca}$ = 1.315 and $Q_{abs}$ = 1.544. **C:** The contours projected to *n-k* space and their intersections identified.

Once solutions are found, forward Mie calculations are performed, the results are compared to the input values, and their relative errors, which we denote Err(*Q*), are computed as

$$\text{Err}(Q_{candidate}) = 100\% \times \left| \frac{Q_{candidate} - Q_{input}}{Q_{input}} \right| \tag{5}$$

Solutions are displayed on the output graph along with the relative errors in both scattering and absorption.

### 3.2 Constraining the inverse problem

In section 3.1, when only $Q_{sca}$ and $Q_{abs}$ are specified, we see that multiple valid solutions of the inversion may exist. Any consideration of the inverse problem would be incomplete without attempting to seek all solutions and evaluate them for physical meaning. For many laboratory experiments, this may be the only practical approach. However, if an additional measured parameter independent of $Q_{sca}$ and $Q_{abs}$ is available, then the solution may be constrained. A favorable choice is to measure the backscatter efficiency $Q_{back}$ with a capable nephelometer (e.g., the TSI Integrating Nephelometer 3563 or Ecotech Aurora 3000). Our example particle from the previous section was found to have $Q_{back}$ = 0.201. We calculate the $Q_{back}$ surface and project the desired contour onto the *n-k* plane (Figure 2A, with $Q_{back}$ truncated above 3.0) and plot all three parameters together to find a unique intersection of the contours at m = 1.77+0.63$i$, shown in Figure 2B. In PyMieScatt, this constraint technique is automatically applied when the user specifies $Q_{back}$ in the function call.

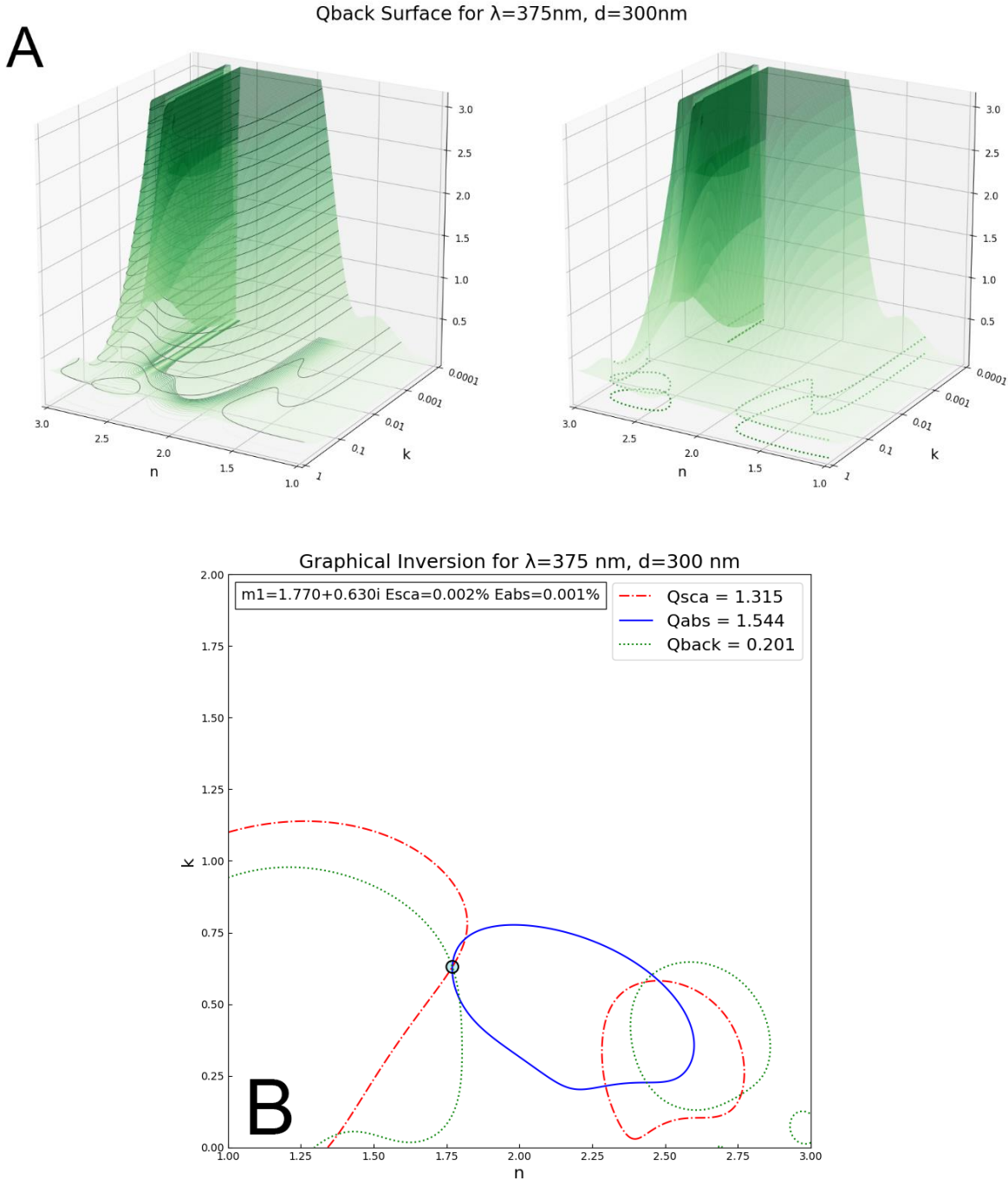

**Figure 2. A:** The $Q_{back}$ surface for a range of *m* with λ=375 nm and d=300 nm (left), and the contours where $Q_{back}$=0.201 (right). **B:** $Q_{sca}$, $Q_{abs}$, and $Q_{back}$ contours projected to *n-k* space and their unique intersection identified at *m*=1.77+0.63*i*. Esca and Eabs follow from Eq. 5 and denote the absolute errors in the $Q_{sca}$ and $Q_{abs}$ values produced by forward calculations using the retrieved refractive index vs. the input values of $Q_{sca}$ and $Q_{abs}$.

In laboratory studies, it is unlikely that instruments will measure particle properties so perfectly, and when considering errors associated with aerosol light scattering and absorption measurements (which are commonly given as a mean and standard deviation over a given averaging time), the refractive index will have associated error as well. Therefore, the solution in *n-k* space is best represented as a region rather than a point. The visualization of these error regions is discussed in section 4.

### 3.3 The survey-iteration method

About 20-50% of the computing overhead for the contour intersection method is devoted to the plotting library. If a non-visual approach is needed by a user (i.e., to include a retrieval of *m* in existing programs), a faster, strictly numerical version of the contour intersection algorithm is included in PyMieScatt, which we call the survey-iteration method. We present an overview of the method here, and complete details may be found in the SM.

This inversion method is a brute-force, error-minimization technique. Inversion by iteration is a downhill-only algorithm that gravitates toward the solution nearest its initial guess, and it seeks local minimum values of error per Eq. 5. This approach estimates the number of solutions for a non-unique problem by first doing a coarse, rapid survey of the scattering and absorption parameter spaces, which is essentially a restrained, low-resolution non-geometric version of the contour intersection method.

From this, the number of distinct solutions is estimated and an initial guess of *m* is generated for each intersection. The algorithm begins at an initial guess and iterates through *n-k* space using small changes of *n* and *k* separately, calculating $Q_{sca}$ and $Q_{abs}$ at each step. The calculated efficiencies are compared to the input values, and Eq. 5 computes $\text{Err}(Q_{i,s})$, where $Q_{i,s}$ is the calculated efficiency (either scattering or absorption) at step *s*. As error decreases, step sizes $\Delta n$ and $\Delta k$ change depending on the magnitude of the relative error. When the calculations are close to the correct value, step *s* will overshoot, error will increase, and the difference

$$\text{Err}(Q_{i,s-1}) - \text{Err}(Q_{i,s}) \qquad (6)$$

becomes negative. When this happens, the routine changes direction (by multiplying either $\Delta n$ or $\Delta k$ by -1), and continues stepping. Once four such sign changes occur, the routine reduces the step size and continues searching. Since $Q_{sca}$ and $Q_{abs}$ are both dependent on *n* and *k*, the algorithm alternates between varying *n* and *k* several times, with the final iteration using the smallest specified step sizes. This process is repeated for each initial guess determined from the survey, and the output from each is the derived *m* as well as the associated errors per Eq. 5. This method is faster than the contour intersection method, though its primary drawback is the handling of the errors associated with the input values. Since the survey-iteration method is strictly numerical, it cannot accurately predict the geometry of the allowed solution region discussed in section 4.

## 3.4 Comparisons and timing

We performed calculations on randomly-generated systems of wavelength and particle size using the contour intersection method and the survey-iteration method to compare timing and robustness. Table 1, while not an exhaustive comparison, shows two examples and illustrates the difference in computing overhead required.

**Table 1.** Examples of solutions found and time required by each algorithm when analyzing a random system.

| Method | Results | Solutions found | Run time (s) |
|---|---|---|---|
| Single Particle | Contour intersection | 5 | 2.343 |
| | Survey-iteration | 5 | 1.485 |
| Size Distribution | Contour intersection | 1 | 81.660 |
| | Survey-iteration | 1 | 38.316 |

## 4 Applications

### 4.1 Refractive index of atmospheric aerosols and handling of measurement errors

Recently, a study conducted at the authors' laboratory investigated the effects of photooxidation on the absorption properties of light absorbing organic aerosol, or brown carbon (BrC) [20]. A pre-release version of PyMieScatt was used to determine the effective refractive indices of BrC particles using $\beta_{sca}$ and $\beta_{abs}$ measurements from integrated photoacoustic-nephelometer spectrometers at 375, 405, and 532 nm, and size distribution measurements from an SMPS. Electron microscopy revealed their morphology to be spherical and homogeneous. Determination of $m$ partially motivated the development of PyMieScatt.

Figure 3 follows the same procedure from section 3.1. In panel A, we show the $\beta_{sca}$ and $\beta_{abs}$ surfaces. In panel B, we locate the contours of 375 nm scattering and absorption measurements of $\beta_{sca} = 5087.2 \pm 361.7$ Mm$^{-1}$ and $\beta_{abs} = 858.3 \pm 46.0$ Mm$^{-1}$, and project those contours to the $n$-$k$ plane. In panel C, we locate their intersection and find the effective $m = 1.576+0.029i$. As mentioned in section 3.2, laboratory measurements have an associated uncertainty. This is often reported as the standard deviation of a time-averaged signal, or in the case of physical measurement, the degree of spread in a transfer function. This uncertainty propagates to the retrieval of $m$. PyMieScatt assumes the physical measurements are accurate and treats errors in optical measurements by visually identifying regions of allowed $m$ as patches of red (indicating the error associated with scattering) and blue (absorption). Where the regions overlap are areas of allowed $m$, where valid solutions within measurement error may exist. This is a strength of the visual method, since not only can we see where solutions may be found, we can also see where they may not.

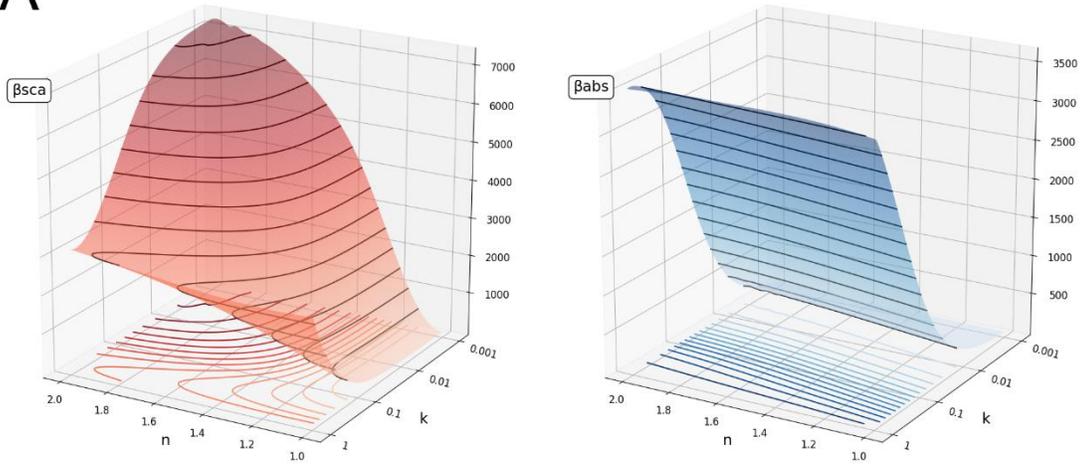

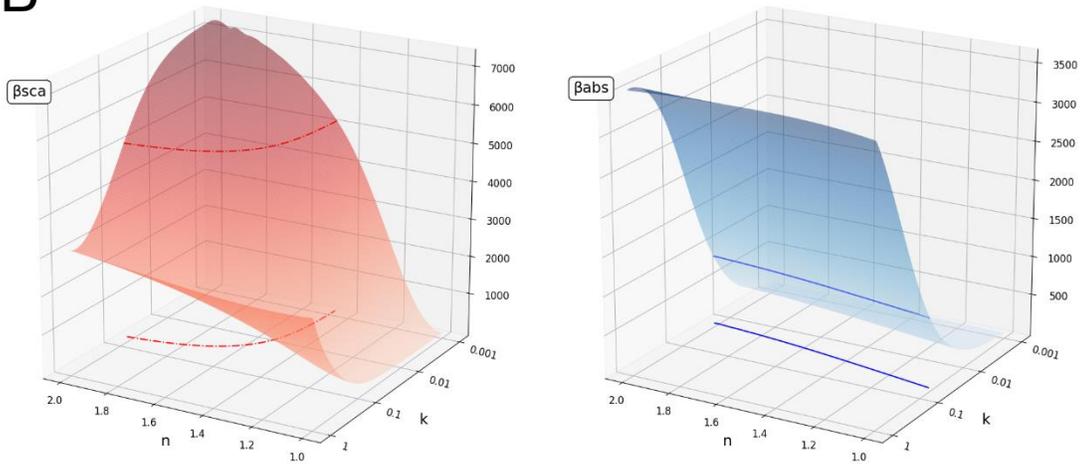

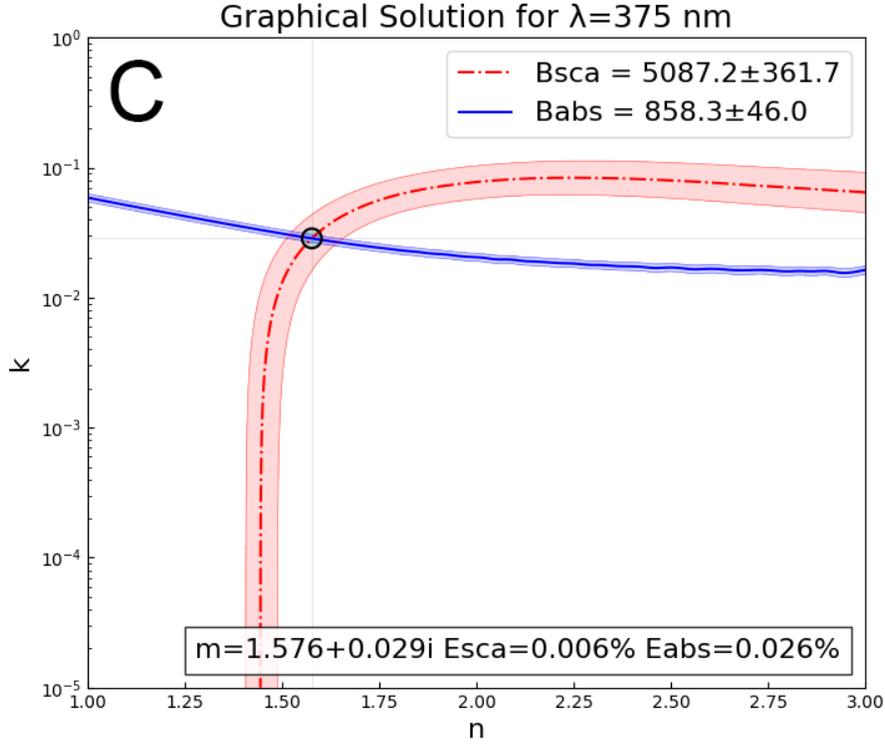

**Figure 3. A:** The $\beta_{sca}$ (left, red) and $\beta_{abs}$ (right, blue) surfaces for a range of $m$ for a size distribution illuminated at $\lambda$=375 nm. **B:** The contours of measured $\beta_{sca}$ = 5087.2 ± 631.7 Mm$^{-1}$ and $\beta_{abs}$ = 858.3 ± 46.0 Mm$^{-1}$. **C:** The contours projected to $n$-$k$ space and their intersection identified at $m = 1.576 + 0.029i$. The shaded regions denote the errors associated with measurements. The overlapping red and blue regions indicate where a valid solution within measurement error may exist. Esca and Eabs follow from Eq. 5 and denote the absolute errors in the $\beta_{sca}$ and $\beta_{abs}$ values produced by forward calculations using the retrieved refractive index vs. the input values of $\beta_{sca}$ and $\beta_{abs}$.

We have shown in section 3.1 that it is possible for Mie theory inversions to have multiple solutions when only two inputs are given, and it is reasonable to ask whether inversions in laboratory studies are yielding the "correct" solution. Here, we show that a distribution of aerosol is likely to have a single physically valid solution, without the need for an additional measured parameter. For atmospheric aerosol, $n$ and $k$ have been observed in the range of 1.3 to 2.0 and $10^{-9}$ to 1.0, respectively [21]. For the sake of illustration, we will extend the analysis of the previous example and evaluate $n$ in the range of 1.0 to 20 while constraining $k$ to between $10^{-3}$ and 1.0. The result is given in Figure 4. The measurement error regions begin to overlap around $n = 5$ and major contours intersect several times at $m$ = 8.417+0.027$i$, 8.840+0.025$i$, 8.941+0.026$i$, 9.592+0.027$i$, and 9.897+0.027$i$. However, only the solution at $m$ = 1.576+0.029$i$ can be considered valid based on our knowledge of the particles in question. This solution is not unique in the mathematical sense, but is the only one that is consistent with physical intuition. We therefore see that $\beta_{back}$ may not be required to obtain a valid, constrained calculation of $m$ for a polydisperse ensemble of particles.

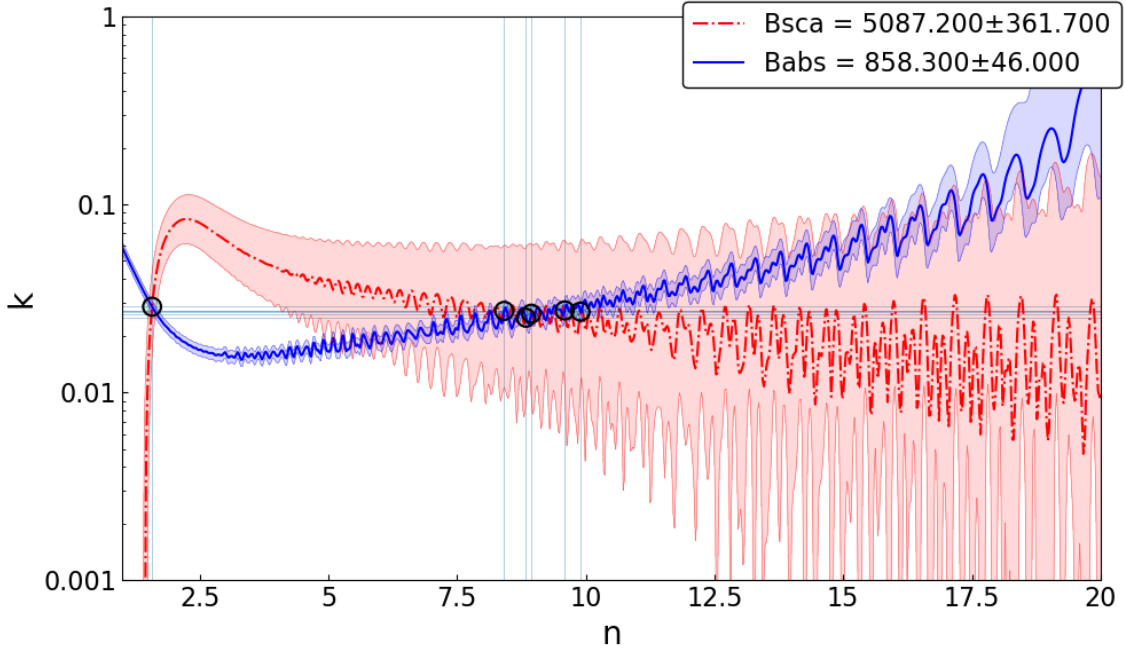

**Figure 4.** The contour intersection inversion method applied to an extended range of *n* from 1.0 to 20.0. The valid solution occurs at *m* = 1.576 + 0.023*i* as before, though several extraneous solutions appear between *n* = 8 and 10. The shaded regions denote the errors associated with measurements, and we have removed some annotations for clarity.

### 4.2 The vanishing degenerate solutions

We have demonstrated that with measurements of only $Q_{abs}$ and $Q_{sca}$, the inverse Mie problem for single particles may have multiple solutions, while for a size distribution and measurements of $\beta_{abs}$ and $\beta_{sca}$, the realistic solution is likely unique (within reason for atmospheric aerosol). Extraneous solutions vanish when we move from a single particle to an ensemble. It is natural to wonder what mechanism governs this transition. PyMieScatt's visualization capabilities were used to study this phenomenon.

Consider a δ-distribution of $10^6$ particles cm$^{-3}$, each 300 nm in diameter, with *m* = 1.60+0.36*i*. This δ-distribution can be considered the limiting case of a lognormal distribution with $d_{pg}$ = 300 and $\sigma_g$ = 1. As $\sigma_g$ increases from 1.0 to 2.0, the distribution behaves like a typical ensemble of aerosols. We plotted the $\beta_{sca}$ and $\beta_{abs}$ surfaces at several values of increasing $\sigma_g$, along with the particle size distribution and the solution for *m* in *n*-*k* space produced by the contour intersection inversion method. We note that the shape of the $\beta_{abs}$ surface remains mostly constant throughout, while the $\beta_{sca}$ surface undergoes the most change. We see the local minimum around *m* = 2.60+0.40*i* lift and flatten as $\sigma_g$ increases. By $\sigma_g \approx$ 1.177, the global minimum responsible for the extra solutions has already lifted above the measurement contours. In this case, it is the scattering behavior of the system that dictates the transition to a unique solution. Other systems may behave similarly.

We have prepared an animation of this phenomenon which can be found in the online supplement to this article. Figure 5 shows several frames of this animation giving an overview of how extra solutions vanish and the desired solution remains unchanged.

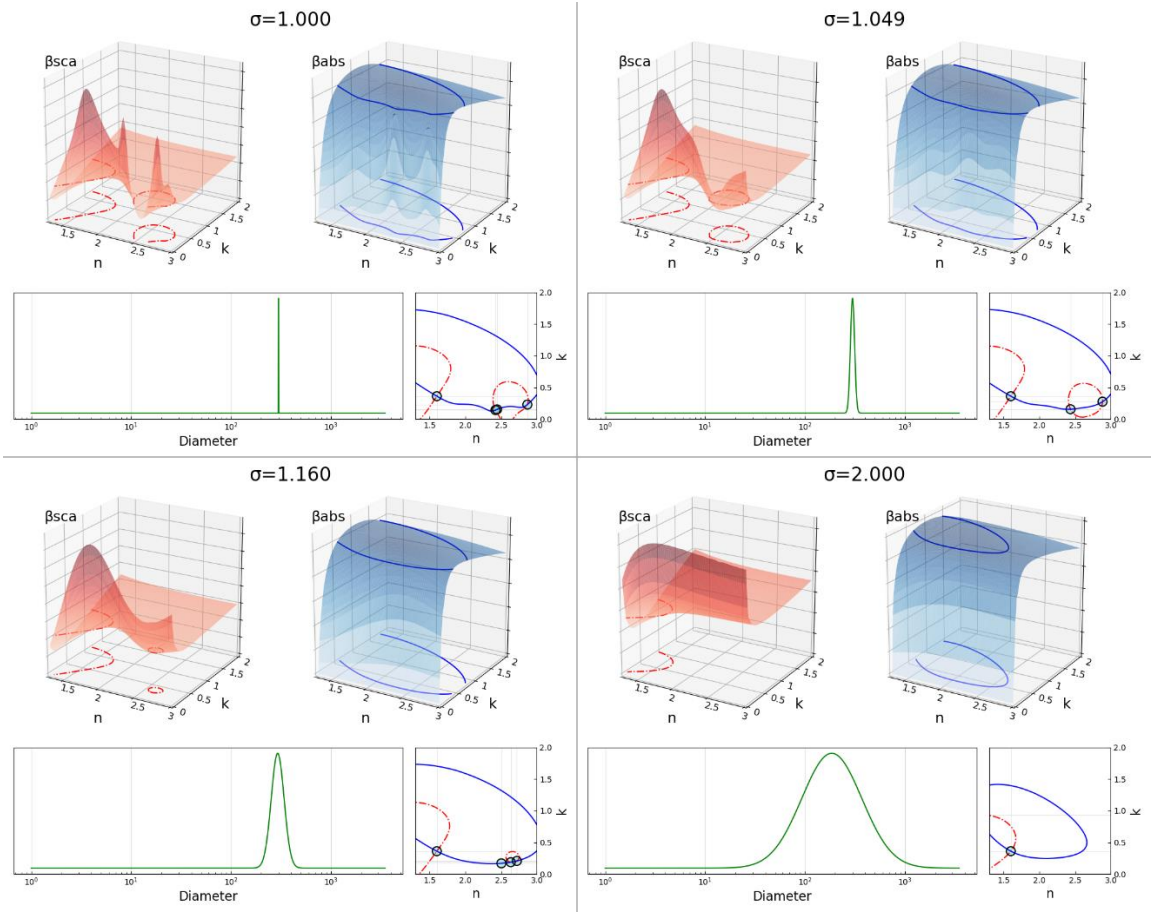

**Figure 5.** The effects on the $\beta_{abs}$ (blue) and $\beta_{sca}$ (red) contours in *n-k* space as $\sigma_g$ increases.

## 5. Concluding Remarks

We conclude by briefly comparing PyMieScatt to other Python implementations of Mie theory. Mie theory has previously been implemented in Python, but not to the level of sophistication of our libraries. Table 2 summarizes the features of several Mie codes available online. In addition to the Mie inversions, PyMieScatt contains over twenty functions for Mie efficiencies of homogeneous particles and core-shell particles; wavelength, diameter, and size parameter ranges; angular scattering functions for single or size distributions of homogeneous particles, as well as single core-shell particles; and functions for arbitrary size distributions, including mathematically-generated k-modal lognormal distributions.

**Table 2.** Features of Python Mie codes found online.

| Feature | py-mie | Pymiecoated | bhmie.py | **PyMieScatt** |
|---|---|---|---|---|
| Homogeneous Sphere | ✓† | ✓†† | ✓† | ✓‡ |
| Coated Sphere | ✓† | ✓†† | | ✓‡ |
| Lognormal Distribution | unimodal | | | k-modal |
| Arbitrary Distribution | | | | ✓ |
| Parameter Ranges | | | | ✓ |
| Angular Functions | | ✓* | ✓** | ✓*** |
| Single-particle Inversion | | | | ✓ |
| Distribution Inversion | | | | ✓ |
| Documentation | limited | | | full |
| Examples | | one | | several |
| Active Maintenance | | | | ✓ |

†Scattering, absorption, and asymmetry parameter
††Scattering, absorption, and backscatter efficiencies; asymmetry parameter; backscatter ratio
‡Extinction, scattering, absorption, and backscatter efficiencies; asymmetry parameter; backscatter ratio; radiation pressure
*only $S_1$ and $S_2$ scalars as functions of a single scattering angle
**$S_1$ and $S_2$ arrays over range of all scattering angles
***$S_1$ and $S_2$ arrays over range of all or user-defined scattering angles; scattering intensity functions for polarized and unpolarized light

PyMieScatt has been applied successfully to theoretical and experimental problems. Our contour intersection inversion method provides visual insight to the problem and aids in identifying multiple solutions and evaluating them for physical validity. Using this method, we have shown how inversions on polydisperse size distributions are unlikely to encounter the issue of multiple solutions. PyMieScatt's code base is open-source, documented, and actively maintained. It was originally developed to aid in very specific calculations to improve analysis of carbonaceous aerosol optical properties, but has evolved to be a general and comprehensive set of tools.

### Data availability
Source code for PyMieScatt is available at https://github.com/bsumlin/PyMieScatt. Details on the experimental setup used for the example in section 4.1 are available in Ref. 20, and the raw data is available from the authors upon request.

### Funding

This work was supported by the NSF under Grant Nos. AGS-1455215 and CBET-1511964, and NASA ROSES under Grant No. NNX15AI66G.

**Supplementary Material for**

# Retrieving the Aerosol Complex Refractive Index using PyMieScatt: A Mie Computational Package with Visualization Capabilities

Benjamin J. Sumlin[1], William R. Heinson[1], and Rajan K. Chakrabarty[1,2]*

[1] Center for Aerosol Science and Engineering

Department of Energy, Environmental and Chemical Engineering

Washington University in St. Louis, Missouri, USA – 63130

[2] McDonnell Center for the Space Sciences

Washington University in St. Louis, Missouri, USA – 63130

* Correspondence to: chakrabarty@wustl.edu

*Authors' note: the SM uses some Python-specific terminology. A glossary of Python terms can be found at https://docs.python.org/3/glossary.html. The authors are not affiliated with this site in any way.*

## 1 A brief description of PyMieScatt

PyMieScatt is an open-source library of Lorenz-Mie (Mie hereafter) theory functions written for Python 3. Source code is available at the Python Package Index (http://pypi.python.org/pypi/PyMieScatt) and GitHub (https://github.com/bsumlin/PyMieScatt/), and can be installed via "pip," a common python package installer. Complete documentation is available at http://pymiescatt.readthedocs.io/. PyMieScatt depends on up-to-date installations of NumPy, SciPy, and Matplotlib, popular modules for mathematical and scientific computing and data visualization [1-3].

Many Mie codes take wavelength and particle size information in the well-known form of the size parameter $x$, a dimensionless quantity expressed as $x = \pi d/\lambda$, where $d$ is the particle diameter and $\lambda$ is the wavelength of incident light. Although we use $x$ in our discussion, we have written our user-facing functions in PyMieScatt to take inputs of $d$ and $\lambda$, since these quantities usually come from separate instruments in experimental setups and we wish to provide a straightforward interface to the mathematics. This section details the equations used in PyMieScatt's Mie functions. A thorough discussion and derivation is given in Bohren and Huffman [1].

### 1.1 Functions for homogeneous spheres

Mie calculations depend on determination of the Mie coefficients $a_n$ and $b_n$ for the scattered electric field amplitudes. These are given by B&H equations 4.56-57, though a more computationally-efficient method is given by equations 4.88-89:

$$a_n = \frac{\left[\frac{D_n(mx)}{m} + \frac{n}{x}\right]\psi_n(x) - \psi_{n-1}(x)}{\left[\frac{D_n(mx)}{m} + \frac{n}{x}\right]\xi_n(x) - \xi_{n-1}(x)} \tag{S1a}$$

$$b_n = \frac{\left[mD_n(mx) + \frac{n}{x}\right]\psi_n(x) - \psi_{n-1}(x)}{\left[mD_n(mx) + \frac{n}{x}\right]\xi_n(x) - \xi_{n-1}(x)} \tag{S1b}$$

where $n$ is an integer that indexes an infinite series and not the real component of $m$. The functions $D_n$, $\psi_n$, and $\xi_n$ are calculated from recurrence relations and remove the need to compute derivatives. Analytically, the subscripts $n$ go to infinity, but in practical application these series may be truncated after $n_{max} = x + 4x^{1/3} + 2$ terms [4]. From $x$, $a_n$ and $b_n$, Mie efficiencies for extinction, scattering, absorption, backscatter, radiation pressure, and asymmetry parameter can be calculated.

Using inputs of *m* and *x* (as separate inputs of particle diameter and wavelength of light), the function MieQ() calculates efficiencies and the asymmetry parameter as follows:

$$Q_{ext} = \frac{2}{x^2} \sum_{n=1}^{n_{max}} (2n+1) \operatorname{Re}\{a_n + b_n\} \quad \text{(S2a)}$$

$$Q_{sca} = \frac{2}{x^2} \sum_{n=1}^{n_{max}} (2n+1) (|a_n|^2 + |b_n|^2) \quad \text{(S2b)}$$

$$Q_{abs} = Q_{ext} - Q_{sca} \quad \text{(S2c)}$$

$$Q_{back} = \frac{1}{x^2} \left| \sum_{n=1}^{n_{max}} (2n+1)(-1)^n (a_n - b_n) \right|^2 \quad \text{(S2d)}$$

$$Q_{ratio} = \frac{Q_{back}}{Q_{sca}} \quad \text{(S2e)}$$

$$g = \frac{4}{Q_{sca} x^2} \left[ \sum_{n=1}^{n_{max}} \frac{n(n+2)}{n+1} \operatorname{Re}\{a_n a_{n+1}^* + b_n b_{n+1}^*\} + \sum_{n=1}^{n_{max}} \frac{2n+1}{n(n+1)} \operatorname{Re}\{a_n b_n^*\} \right] \quad \text{(S2f)}$$

$$Q_{pr} = Q_{ext} - g Q_{sca} \quad \text{(S2g)}$$

For particles that are small compared to the wavelength ($x = \pi d/\lambda \ll 1$), in what is usually called the Rayleigh or low-frequency limit), the functions RayleighMieQ() and LowFrequencyMieQ() can compute efficiencies faster by appealing to some simplifying assumptions. For sufficiently small *x*, RayleighMieQ() can be used to compute efficiencies by:

$$Q_{sca} = \frac{8x^4}{3} \left| \frac{m^2 - 1}{m^2 + 2} \right|^2 \quad \text{(S3a)}$$

$$Q_{abs} = 4x \operatorname{Im}\left\{ \frac{m^2 - 1}{m^2 + 2} \right\} \quad \text{(S3b)}$$

$$Q_{ext} = Q_{sca} + Q_{abs} \quad \text{(S3c)}$$

$$Q_{back} = \frac{3 Q_{sca}}{2} \quad \text{(S3d)}$$

$$Q_{ratio} = 1.5 \quad \text{(S3e)}$$

$$g = 0 \quad \text{(S3f)}$$

$$Q_{pr} = Q_{ext} \quad \text{(S3g)}$$

LowFrequencyMieQ() uses the same equations as MieQ(), but only computes the first few terms of $a_n$ and $b_n$:

$$a_1 = \left(\frac{m^2-1}{m^2+2}\right) \times \left[-\frac{i2x^3}{3} - \frac{i2x^5}{5}\frac{(m^2-2)}{(m^2+2)} + \frac{4x^6}{9}\left(\frac{m^2-1}{m^2+2}\right)\right] \tag{S4a}$$

$$a_2 = -\frac{ix^5}{15}\frac{(m^2-1)}{2m^2+3} \tag{S4b}$$

$$b_1 = -\frac{ix^5}{45}(m^2-1) \tag{S4c}$$

$$b_2 = 0 \tag{S4d}$$

For particles that are small compared to the wavelength ($x \ll 1$, in what is usually called the Rayleigh or low-frequency limit), computation of efficiencies can be made significantly faster by appealing to some simplifying assumptions. Benchmarks for timing of various functions may be found in section 3 of this document.

### 1.2 Functions for size distributions of homogeneous spheres

Laboratory and field experiments often involve a polydisperse distribution of particles. Optical measurements are integrated over these distributions, and the quantities reported are the coefficients $\beta$ rather than efficiencies $Q$. We expedite this integration with self-contained functions Mie_SD(), which takes binned size distribution data such as one would measure with a scanning mobility particle sizer (SMPS), and Mie_Lognormal(), which mathematically generates a k-modal lognormal distribution by specifying the geometric means ($d_{pg}$), geometric standard deviations ($\sigma_g$), number of particles ($N_\infty$), and proportionality constants that allocate some fraction of $N_\infty$ to each mode. However convenient they may be, analytic size distributions do not represent real-world size distributions perfectly. There will always be some deviation, though for prototyping the behavior of a given system, it is widely acknowledged that certain mathematical forms are suitable. The study of arbitrary distributions is important to aerosol physics, and we have written PyMieScatt to allow the user to construct any particle size distribution (or acquire one from laboratory experiments) and study its optical properties.

The analytical expression for the coefficients is

$$\beta_i = (10^{-6}) \int_0^\infty \frac{\pi d_p^2}{4} Q_i(m,\lambda,d_p) n_d(d_p) \mathrm{d}d_p \tag{S5a}$$

where $d_p$ (in nm) is the diameter of the particle, $n_d(d_p)$ (in cm$^{-3}$) is the number density of particles of diameter $d_p$, and the subscript $i$ represents an optical parameter (absorption, scattering, etc.). The units on each quantity in Eq. 2 are typical of units reported by instrumentation such as an SMPS, and we chose to keep the inputs consistent with laboratory measurements. The factor $10^{-6}$ is a multiplicative factor used to change the units of the equation to Mm$^{-1}$, the common unit for extinction measurements.

The conversion factor is derived as follows. For the moment, neglect the multiplicative factor.

$$\beta = \int_0^\infty \frac{\pi d_p^2}{4} Q_i(m, \lambda, d_p) n_d(d_p) \mathrm{d}d_p \tag{S5b}$$

The units on each component inside the integral are $d_p$ in nm, and $n_d(d_p)\mathrm{d}d_p$ in cm$^{-3}$. To convert to Mm$^{-1}$:

$$\left(10^{-9} \frac{\mathrm{m}}{\mathrm{nm}}\right)^2 \left(10^2 \frac{\mathrm{cm}}{\mathrm{m}}\right)^3 \left(10^6 \frac{\mathrm{m}}{\mathrm{Mm}}\right) = 10^{-6} \tag{S5c}$$

Therefore, a factor of $10^{-6}$ cm$^3$ nm$^{-2}$ Mm$^{-1}$ is needed to properly convert the units.

The bulk asymmetry parameter $G$ is calculated by the weighted average of asymmetry parameters of individual diameters $g(d_p)$ [6, 7]:

$$G = \frac{\int g(d_p) \beta_{sca}(d_p) \mathrm{d}d_p}{\int \beta_{sca}(d_p) \, \mathrm{d}d_p} \tag{S6}$$

### 1.3 Scattering matrix elements and the scattered field

PyMieScatt can produce output suitable for plotting, for example, the angular scattering intensity. Output can be constrained to an angular range with a specified angular resolution, otherwise it defaults to 0-180° in 0.5° increments. Calculations of the scattered field intensity follow from computation of $S_1$ and $S_2$:

$$S_1 = \sum_{n=1}^{n_{max}} \frac{2n+1}{n(n+1)} (a_n \pi_n + b_n \tau_n) \tag{S7a}$$

$$S_2 = \sum_{n=1}^{n_{max}} \frac{2n+1}{n(n+1)} (a_n \tau_n + b_n \pi_n) \tag{S7b}$$

where $S_1$, $S_2$, $\pi_n$ and $\tau_n$ are all functions of scattering angle. The functions $\pi_n$ and $\tau_n$ are calculated from recurrence relations. The parallel (*SR*), perpendicular (*SL*), and unpolarized (*SU*) intensities can be calculated, along with the four nonzero scattering matrix elements $S_{11}$, $S_{12}$, $S_{33}$, and $S_{34}$ (see Eq. S8 and S9). PyMieScatt can also compute the angular scattering intensity of a distribution of particles in either $\theta$-space or q-space by integrating over the size distribution.

In addition to homogeneous spheres, PyMieScatt includes functions to calculate efficiencies, scattering intensity functions, and matrix elements for coated spheres, given the diameters and refractive indices of both the core and shell. The computation of $a_n$ and $b_n$ for coated spheres closely follows the original BHMIE.

If the angular scattering intensity is needed, an array of values can be produced. Output can be constrained to an angular range with a specified angular resolution, otherwise it defaults to 0-180° in 0.5° increments. Calculations of the scattered field intensity follow from computation of $S_1$ and $S_2$, which are given by equations 4a and 4b in the main text. The parallel (*SR*), perpendicular (*SL*), and unpolarized (*SU*) intensities are calculated by

$$SR(\theta) = |S_1|^2 \tag{S8a}$$

$$SL(\theta) = |S_2|^2 \tag{S8b}$$

$$SU(\theta) = \frac{1}{2}(SR + SL) \tag{S8c}$$

Additionally, from the $S_1$ and $S_2$ parameters, we can calculate the four nonzero scattering matrix elements $S_{11}$, $S_{12}$, $S_{33}$, and $S_{34}$ using the MatrixElements() function:

$$S_{11} = \frac{1}{2}(|S_2|^2 + |S_1|^2) \tag{S9a}$$

$$S_{12} = \frac{1}{2}(|S_2|^2 - |S_1|^2) \tag{S9b}$$

$$S_{33} = \frac{1}{2}(S_2^*S_1 + S_2 S_1^*) \tag{S9c}$$

$$S_{34} = \frac{i}{2}(S_1 S_2^* - S_2 S_1^*) \tag{S9d}$$

## 2 Details of the survey-iteration inversion algorithm

The survey-iteration inversion algorithm is a brute-force guess-and-check algorithm that minimizes error in retrieved *m* based on forward Mie calculations. In the survey phase, coarse arrays of $Q_{sca}(n,k)$ and $Q_{abs}(n,k)$ (or $\beta_{sca}(n,k)$ and $\beta_{abs}(n,k)$) are constructed and surveyed for values close to the input efficiencies (or coefficients). Array locations with indices that are common to both arrays are considered candidate solutions. There may be cases where this survey over-predicts the number of distinct solutions since multiple locations may be within range of an actual solution. However, the iteration phase will, from these multiple candidates, produce a distinct solution in a given neighborhood. Each neighborhood is small enough that the correct number of distinct solutions can be predicted without user adjustment, however, in cases where the contour intersection and survey-iteration method disagree, the survey resolution can be refined by optional user input.

The iteration phase is best described by the flowchart given in Figure S1. The real part of the refractive index is treated first in the red "Scattering" loop, and then the imaginary part is treated by the blue "Absorption" loop. The algorithm runs for each candidate *m* neighborhood found by the survey.

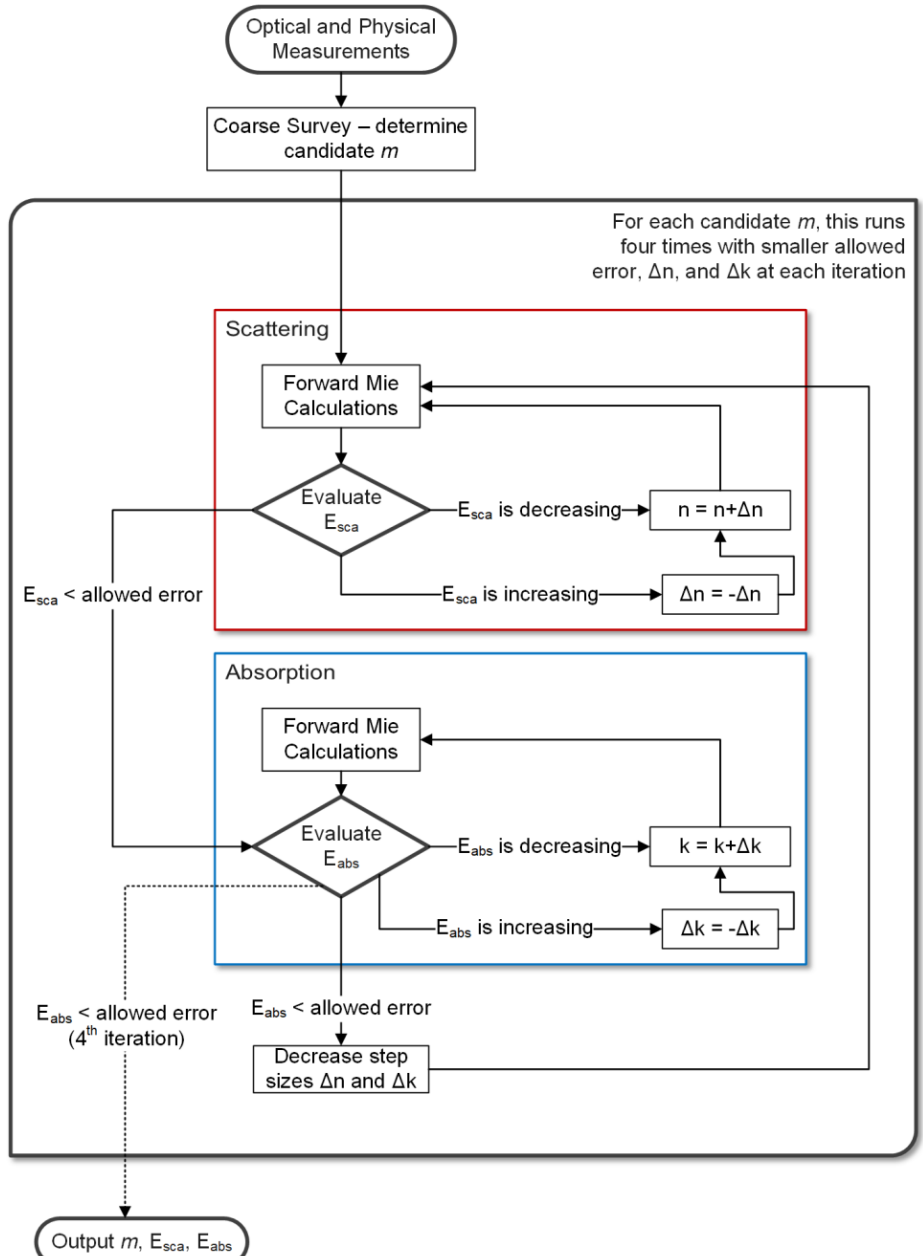

**Figure S1.** Flowchart describing the iteration phase of the survey-iteration algorithm. $E_{abs}$ and $E_{sca}$ are the errors in scattering and absorption efficiencies (or coefficients) given by Eq. 5 of the main text.

## 3 Benchmarks for selected functions

Python includes simple timing libraries that facilitate benchmarking. The following tests were performed on a personal computer with an Intel Core i5-6300U (2.4 GHz clock speed) and 8 GB of RAM, typical of a modern personal computer. Three functions were tested: MieQ(), RayleighMieQ(), and Mie_Lognormal(). MieQ() calculates efficiencies directly using Mie theory, while RayleighMieQ() simplifies the calculations for particles small compared to the wavelength ($x = \pi d/\lambda \ll 1$), and Mie_Lognormal() computes the Mie coefficients, which are efficiencies integrated over a size distribution.

For the tests, inputs are completely randomized across all input parameters. MieQ() and RayleighMieQ() were performed with $10^5$ separate sets of randomized input parameters, while Mie_Lognormal() was given 100 separate randomized size distributions of $10^5$ particles. The code used for benchmarking was:

```
>>> import PyMieScatt as ps
>>> import numpy as np
>>> from time import time
>>>
>>> d = np.random.uniform(100,1000,100000)
>>> w = np.random.uniform(200,1000,100000)
>>> n = np.random.uniform(1,3,100000)
>>> k = np.random.uniform(0.0001,1,100000)
>>> b = time()
>>> for dd,ww,nn,kk in zip(d,w,n,k):
>>>     _q = ps.MieQ(nn+1j*kk,ww,dd)
>>> e = time()
>>> print("Mie:\t{s:1.2f} seconds.".format(s=e-b))
>>>
>>> d = np.random.uniform(10,100,100000)
>>> w = np.random.uniform(200,500,100000)
>>> n = np.random.uniform(1,3,100000)
>>> k = np.random.uniform(0.0001,1,100000)
>>> b = time()
>>> for dd,ww,nn,kk in zip(d,w,n,k):
>>>     _q = ps.RayleighMieQ(nn+1j*kk,ww,dd)
>>> e = time()
>>> print("Rayleigh:\t{s:1.2f} seconds.".format(s=e-b))
>>>
>>> sigmag = np.random.uniform(1.1,2,100)
>>> dpg = np.random.uniform(200,500,100)
>>> w = np.random.uniform(200,500,100)
>>> n = np.random.uniform(1,3,100)
>>> k = np.random.uniform(0.0001,1,100)
>>> b = time()
>>> for ss,d,ww,nn,kk in zip(sigmag,dpg,w,n,k):
>>>     _q = ps.Mie_Lognormal(nn+1j*kk,ww,ss,d,1e5,upper=2000)
>>> e = time()
>>> print("Lognormal:\t{s:1.2f} seconds".format(s=e-b))
```

Over $10^5$ iterations, MieQ() completed in 29.26 s (292.6 µs per iteration) and RayleighMieQ() completed in 1.03 s (10.3 µs per iteration). Over 100 iterations, Mie_Lognormal() completed in 21.13 s (211.3 ms per iteration).

Two additional cases were tested. In the previous test, each calculation was stored to the variable _q, and it was overwritten each time a new calculation was performed. When omitting the need to even temporarily store the result (and all inputs and iteration counts the same), MieQ() and RayleighMieQ() completed $10^5$ iterations in 24.75 and 0.81 s, respectively, and Mie_Lognormal() completed 100 iterations in 18.77 s.

When the results of individual calculations are stored in lists, MieQ() and RayleighMieQ() completed $10^5$ iterations in 26.31 and 0.87 s, respectively, and Mie_Lognormal() completed 100 iterations in 20.65 s. Memory usage is inconsistent due to the way Python allocates memory for lists, and is not reported here.